\documentclass[twocolumn, superscriptaddress]{revtex4-2}
\usepackage{graphicx} 
\usepackage[caption=false]{subfig}
\usepackage{physics}
\usepackage{bm}
\usepackage{xcolor}
\usepackage{enumerate}
\usepackage{booktabs}
\usepackage{multirow}
\usepackage{makecell}
\usepackage{hyperref, soul}
\usepackage[capitalize]{cleveref}
\hypersetup{
    colorlinks,
    linkcolor={blue},
    citecolor={blue},
    urlcolor={blue}
}

\begin{document}

\title{Scalable quantum error correction tailored for a heavy-hex qubit array}

\author{Seok-Hyung Lee}
\affiliation{School of Physics, The University of Sydney, Sydney, New South Wales 2006, Australia}
\affiliation{Department of Quantum Information Engineering, Sungkyunkwan University, Suwon 16419, Republic of Korea}

\author{Xanda C. Kolesnikow}
\affiliation{School of Physics, The University of Sydney, Sydney, New South Wales 2006, Australia}

\author{Jun Zen}
\affiliation{Okinawa Institute of Science and Technology, Okinawa, Japan}

\author{Evan T. Hockings}
\affiliation{School of Physics, The University of Sydney, Sydney, New South Wales 2006, Australia}

\author{Campbell K. McLauchlan}
\affiliation{School of Physics, The University of Sydney, Sydney, New South Wales 2006, Australia}

\author{Georgia M. Nixon}
\affiliation{School of Physics, The University of Sydney, Sydney, New South Wales 2006, Australia}

\author{Thomas R. Scruby}
\affiliation{Okinawa Institute of Science and Technology, Okinawa, Japan}

\author{Stephen D. Bartlett}
\email{stephen.bartlett@sydney.edu.au}
\affiliation{School of Physics, The University of Sydney, Sydney, New South Wales 2006, Australia}

\author{Robin Harper}
\affiliation{School of Physics, The University of Sydney, Sydney, New South Wales 2006, Australia}

\author{Benjamin J. Brown}
\affiliation{IBM Quantum, T. J. Watson Research Center, Yorktown Heights, New York 10598, USA}
\affiliation{IBM Denmark, Sundkrogsgade 11, 2100 Copenhagen, Denmark}

\begin{abstract}
To produce an operable quantum computer that is made with imperfect hardware, we must design and test scalable quantum error correcting codes that are suited for the devices we can build and, in unison, develop decoding strategies that accommodate device-specific noise characteristics.  Here, we introduce the \emph{dynamic compass code}, a subsystem code with a novel syndrome extraction cycle, that has a competitive threshold while making efficient use of qubits arranged on a heavy-hex lattice.  We use a superconducting qubit array to implement a distance-5 instance of this code, and demonstrate how detailed noise characterisation can 
boost decoder performance to yield significant improvements in logical error rates.  We perform averaged circuit eigenvalue sampling (ACES) to acquire detailed context-dependent error information on all elements of the syndrome extraction process. Furthermore, we leverage soft information produced from measurement devices to augment the decoder with measurement error information and detect leakage errors for exclusion through post-selection.  
Our noise-informed approach yields up to 38.3\% improvement in the logical error rate of a distance-5 implementation of the dynamic compass code in experiment.
\end{abstract}

\maketitle

\section{Introduction}

In order to produce high-fidelity, error-corrected logical qubits for fault-tolerant quantum computing, we must design codes with syndrome extraction circuits that are readily implemented with hardware, and we must tailor our decoding strategies to optimise performance under real-world and device-dependent conditions. In the long term these endeavours will minimise the physical hardware resource requirements needed to complete a fault-tolerant algorithm with some target logical error rate. In the near term, progress to this end will give rise to experimental demonstrations of high-fidelity logical qubits.

Here we introduce the {\it dynamic compass code}: a code that is suited for implementation on qubits arranged on a two-dimensional heavy-hex lattice, and that possesses a syndrome extraction circuit composed of a dynamic sequence~\cite{Hastings2021dynamically, gidney2023baconthreshold, alam2025baconshorboardgames} of weight-two and -four checks. Unlike the heavy-hex code~\cite{chamberland2020} (its subsystem code precursor), our dynamic code demonstrates a threshold~\cite{DCCnumerics}, and thus provides a route to scale to arbitrarily small logical error rates given a larger device. Here we interrogate the performance of error-correction strategies in practice by operating a distance-5 instance of the dynamic compass code on an array of superconducting qubits.

Quantum error correction relies on a decoding algorithm that takes syndrome data together with prior information about device noise to attempt to reverse the effects of errors.  While considerable research effort has gone into improving decoders, they have commonly been abstracted from the specific hardware and therefore tend to rely on simplistic assumptions about the device noise.  As such, even optimised decoders may not perform well in practice.  As larger scale quantum devices become available, we can look to improve real-world decoding performance by furnishing the algorithm with data obtained from the device used to implement the code. In this work we make use of recent highly optimised matching decoders~\cite{Higgott2025sparseblossom,higgott2023improved} and demonstrate significant improvements in logical error rates by exploiting detailed noise information. Specifically, we demonstrate reductions in logical error rates by \textbf{24\%--38\%} by supplying these decoders with a combination of detailed, context‑dependent noise characterisation of the device \cite{hockings2024}, soft decoding through analysis of IQ data \cite{pattison2021improvedquantumerrorcorrection,Ali2024,majaniemi2025reducing} and leakage detection \cite{sundaresan2022}.

To begin, we inform our decoder with a tailored noise characterisation of the device acquired through averaged circuit eigenvalue sampling (ACES)~\cite{flammia2021b,hockings2024,hockings2025improvingerrorsuppressionnoiseaware}. 
We find this improves logical error rates by 30.4\% (9.4\%) for the $X$ ($Z$) basis, relative to a simple noise model based on the hardware calibration snapshot.
ACES gives a rapid, 
comprehensive characterisation of the syndrome‑extraction circuits,  obtaining a full, gate‑ and layer‑resolved Pauli error model for all operations in the cycle, including mid‑circuit measurements. This characterisation produces a scalable, high‑fidelity noise model that can be readily integrated into any decoder.

We find a further improvement of an additional 11.3\% (16.3\%) in $X$ ($Z$) basis logical error rate by using soft information from readout of each shot as input for the decoder, augmenting the decoder with measurement error information and detecting leakage for post-selection.
We incorporate soft information from resonator‑based readout by using weighted Gaussian mixture models to transform raw IQ samples into relative probabilities of each measurement outcome. 
Raw IQ data has been previously used with neural network decoders \cite{Varbonov2025}, and transformed data with repetition/bit-flip codes \cite{Ali2024,hanisch2025softinformationdecodingsuperconducting}.
We show how the same IQ data enables real‑time identification of leakage events, providing substantial gains, when fewer than 4\% of runs per round are rejected through post‑selection.

These results provide an experimental demonstration that substantial, scalable, and practically relevant improvements in QEC performance on contemporary superconducting hardware can be achieved by combining full-cycle, device-specific noise information and analog information from measurement devices.
The techniques demonstrated here scale efficiently to larger devices, providing a blueprint for noise‑informed QEC strategies in the era of quantum processors with hundreds of qubits.

\begin{figure*}[t]
    \centering
    \includegraphics{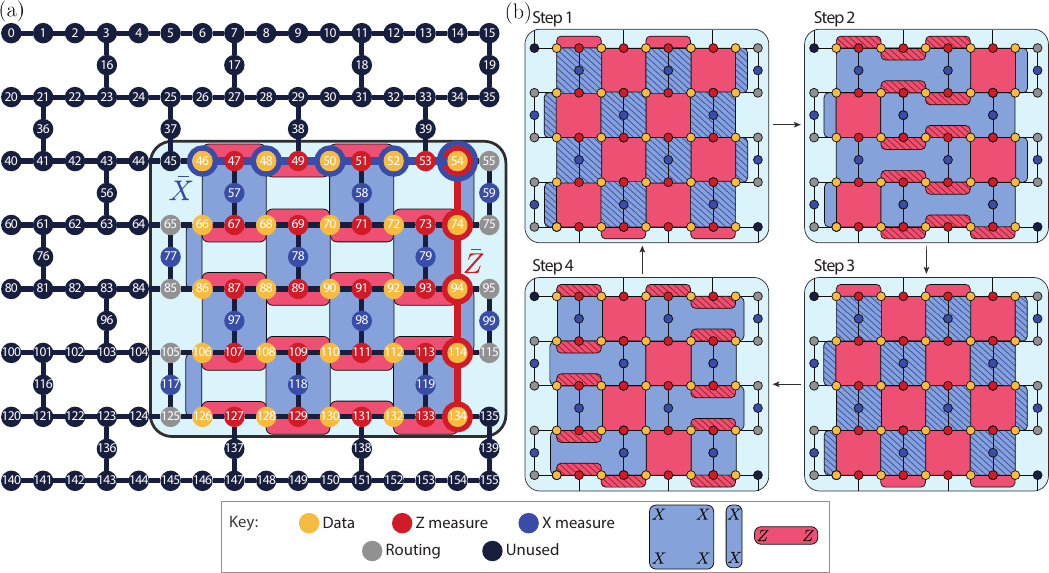}
    \caption{{\bf Dynamic compass code} \textbf{(a)} Three-valent heavy-hex layout of the IBM Quantum 156-qubit Heron revision-3 quantum processor \texttt{ibm\_pittsburgh}. Numbered dots are qubits and lines indicate qubit couplings. The distance-5 heavy-hex experiments are performed in the light blue region.  We label this code placement by Qubit 46, the lowest index value (top-left) qubit used in the arrangement.  An alternative distance-5 code placement based on Qubit 65 is analysed in the Appendix.
    Weight-four $X$-type (weight-two $Z$-type) checks are shown in blue (red) shaded boxes. 
    Logical $\bar{X}$ and $\bar{Z}$ operators for the patch are defined along the blue and red strings respectively. 
     \textbf{(b)} The dynamic compass code schedule with four sub-steps. The dashed boxes show the checks measured in each sub-step. The remaining coloured boxes indicate the current instantaneous stabiliser group, which will commute with the current measured checks. In steps 1 and 3, all $X$-type checks are measured. In steps 2 and 4, a different subset of $Z$-type checks are measured as shown.}
    \label{fig:device_schedule}
\end{figure*}

\section{Results} \label{sec:results}

\subsection{Quantum codes on a heavy-hex qubit array}

We perform experiments using the IBM Quantum \textit{Heron} class of superconducting quantum processor, where 156 qubits are arranged on a plane with connectivity described by a heavy-hex lattice, displayed in \cref{fig:device_schedule}(a).
We present results on the revision-3 device \texttt{ibm\_pittsburgh}, which supports mid-circuit measurements and high fidelity ($\simeq 99.9\%$) two-qubit gates. This generation of quantum processor, with its size, fidelity, and capabilities opens up the possibility for in-depth analysis and characterisation of logical operations on error-corrected qubits.

The heavy-hex code~\cite{chamberland2020} is a variant of the well-studied surface code that is adapted to the connectivity of the heavy-hex lattice.  We place a distance-5 instance of this code on the quantum processor as shown in \cref{fig:device_schedule}(a), where we have used detailed error characterisation and modelling of the device to choose a location that maximises performance following the approach of Ref.~\cite{harper2025characterising}.  We choose a distance-5 code instance in this study for a number of reasons.  While distance-3 code instances have excellent performance at a number of locations on devices of this type~\cite{harper2025characterising}, the simplicity of decoding very small codes such as distance-3 makes it challenging to observe the potential gains from innovative decoder strategies such as the noise-informed approach studied here (see also Ref.~\cite{hockings2025improvingerrorsuppressionnoiseaware}).  On the other hand, while a single placement of a distance-7 code instance is possible on the present device, a small number of under-performing qubits and couplers result in very poor performance of this large code, which would limit our study.  A distance-5 instance proves to be a fertile compromise for our investigations.

While small instances of the heavy-hex code can yield good performance~\cite{sundaresan2022, harper2025characterising}, this code family possesses several deficiencies that make it ill-suited for fault-tolerant quantum computing applications.  These deficiencies stem from the fact that this code family has $X$-type stabilisers that are extensive:  as the distance of the code increases, the weight of these stabilisers grows linearly with the distance.  
Roughly speaking, since each stabiliser is responsible for identifying errors acting on a number of qubits that grows with the code distance, the stabilisers become ineffective for large distance codes.
The result is that this code family does not demonstrate an error threshold, a critical property used in fault-tolerant constructions. Thus, the heavy-hex code will not allow for arbitrary reductions in the logical error rate.  This is also the case for the well-studied Bacon-Shor code~\cite{Bacon2006operator,AliferisCross2007}.

\subsection{Dynamic compass code}

We introduce the \emph{dynamic compass code}, which builds on the heavy-hex code, using a dynamic measurement schedule to overcome its deficiencies.  Recent results have demonstrated that dynamic measurement schedules for the check operators of the Bacon-Shor code~\cite{Bacon2006operator}, where only a subset of checks are measured at each round, can yield families of syndrome extraction circuits that demonstrate thresholds~\cite{gidney2023baconthreshold,alam2025baconshorboardgames}.
Crucially, a measurement schedule can be chosen such that the detectors of the syndrome extraction circuit take up constant space-time volume even as the size of the code is increased~\cite{alam2025baconshorboardgames}. As a result, the detector error rates can be bounded.

A dynamic measurement schedule can be applied to the heavy-hex code, with exceptional results. We refer to the resulting code as the dynamic compass code, owing to its similarity to the (subsystem) 2D compass codes of Refs.~\cite{li2019, chamberland2020}.  We present the measurement schedule in Fig.~\ref{fig:device_schedule}(b). Certain two-body $X$ measurements are absent in steps 2 and 4, which is the crucial part that allows the $Z$ detectors to be maintained at constant weight. The $X$ detectors are therefore formed every two measurement rounds, while $Z$
detectors are formed every round. This means that the time-like distance for some of the $X$ checks is reduced which involves some non-trivial tradeoffs for finite code distances. 

Numerical evidence to support the presence of a threshold for the dynamic compass code is presented in Ref.~\cite{DCCnumerics}, where it is also shown that different choices of schedule can provide tradeoffs in $X$ vs $Z$ performance, as well as threshold for stability experiments.  

Here, we consider a distance-5 instance of the dynamic compass code implemented as shown in Fig.~\ref{fig:device_schedule}, with the same footprint as the heavy-hex code.  We use the same syndrome extraction circuits as for the heavy-hex code, as detailed in Ref.~\cite{harper2025characterising}, with the key difference with the dynamic compass code being the choice of measurement schedule.  The measurements in steps 1 and 2 can be performed simultaneously to reduce idle time, and similarly the measurements in steps 3 and 4, using the compressed circuits proposed in Ref.~\cite{harper2025characterising}.

\begin{figure*}
    \includegraphics[width=\linewidth]{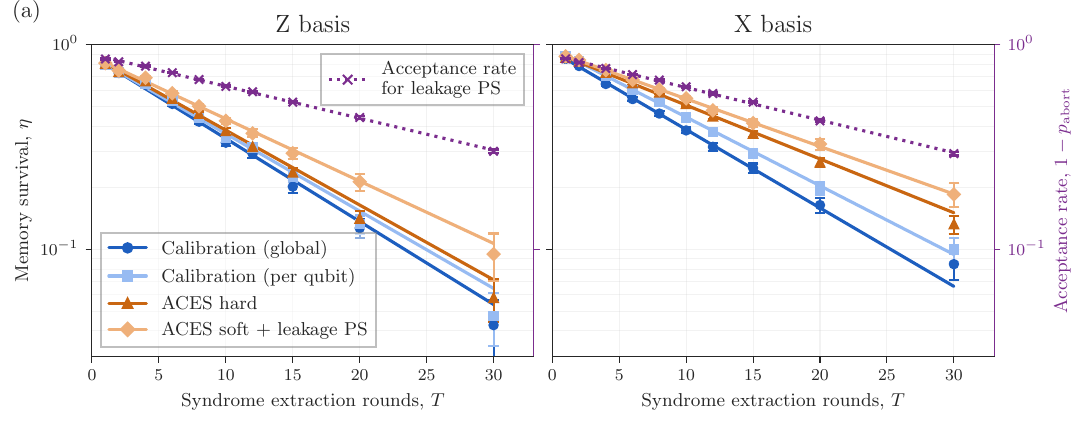} \\
    \includegraphics[width=0.6\linewidth]{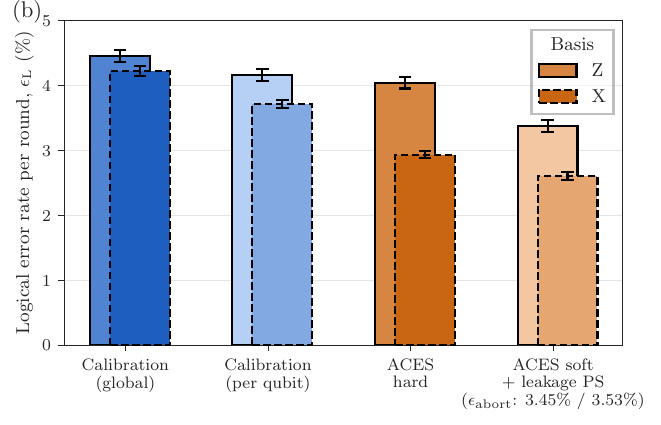}
    \includegraphics[width=0.39\linewidth]{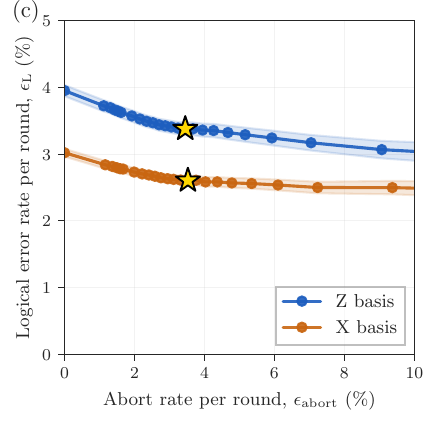}
    \caption{ \textbf{Memory experiment results for the distance-5 dynamic compass code with different decoding strategies.}
        \textbf{(a)}~Memory survival rates plotted against the number of syndrome extraction rounds under different noise-informed decoding methods: two baselines using global (qubit-averaged) or per-qubit IBM calibration data, and our ACES-based methods with/without soft decoding and leakage post-selection.
        For leakage post-selection we plot the abort rates in purple.
        The memory survival rate is defined as $\eta := 1 - 2p_\mathrm{L}$, where $p_\mathrm{L}$ is the logical error rate averaged over two basis states ($\ket{0}$ and $\ket{1}$ for $Z$ basis and $\ket{\pm}$ for $X$ basis).
        We fit $\log\eta$ linearly as a function of $T$, and use the fitted slope $G$ to estimate the logical error rate per round: $\epsilon_\mathrm{L} = (1 - e^G)/2$.
        \textbf{(b)}~Logical error rate per round for the four decoding methods, resulting in the overall relative improvement of 24.2\% (38.3\%) for the $Z$ ($X$) basis.
        For leakage post-selection, we specify the per-round abort rates $\epsilon_\mathrm{abort}$ for the Z and X bases, respectively.
        \textbf{(c)}~Trade-off between per-round logical error rate and per-round abort rate for the ``ACES soft decoding + leakage PS'' setting. 
        Each data point corresponds to a cutoff on the leakage probability estimated from soft information; a sample is aborted if any measurement's leakage probability exceeds this cutoff. 
        Data points at the cutoff of 0.5, used in (a) and (b), are highlighted as yellow stars.
    }
\label{fig:ler_analysis_results}
\end{figure*}

\subsection{Noise-Informed Decoding}

As with the surface code and Bacon-Shor code, the dynamic compass code can be decoded using a matching decoder (i.e., a decoder based on minimum-weight perfect-matching) such as PyMatching \cite{Higgott2025sparseblossom} or BeliefMatching \cite{higgott2023improved}.  
Under a uniform circuit-level noise model—where each unitary gate is preceded by a single- or two-qubit depolarising channel, each idling qubit undergoes a depolarising channel, each reset (measurement) is followed (preceded) by a bit-flip error, and all error rates are set equal—the dynamic compass code exhibits a threshold determined numerically to exceed $10^{-3}$~\cite{DCCnumerics}.

It is important to recall, however, that matching decoders perform near-optimally only if provided with an accurate noise model. 
That is because, given a syndrome, matching decoders aim to identify the error configuration that has the maximum likelihood, which depends on the prior probabilities of individual elementary errors.
A standard assumption for performing simulations on decoder performance is to assume the uniform circuit-level noise model, but if the actual error distribution is not well-described by this simple distribution, the performance of the decoder may be considerably reduced.  

For our distance-5 implementation of the dynamic compass code on the superconducting quantum processor described above, we now consider how to use detailed error characterisation of the quantum processor as well as additional information from the qubit measurement process to inform the decoder and improve performance.

\paragraph*{Error characterisation with ACES.---}
ACES provides an efficient and scalable method of characterising all of the gate operations in a syndrome extraction circuit \cite{hockings2024}, including characterisation of the mid-circuit measurements, determining the errors on each gate in the context of the circuit layer. We use the open-source implementation of ACES~\cite{Hockings_QuantumACES_jl_design_noise_2025} to design and characterise the circuits used to perform syndrome extraction for both the heavy-hex code and the dynamic compass code.  All data required for ACES can be acquired in approximately 30 minutes of QPU time for a choice of code, circuit, and location on the device. 
ACES characterisation outputs the relevant Pauli channel for each gate, in each layer of non‑overlapping gates. For measurements, it provides both bit-flip error probabilities as well as the probability of a classical read-out error. 
These noise characterisation data are used to construct a shot-independent base detector error model 
(see Methods Sec.~\ref{sec:soft_decoding}), which can then be provided to a matching decoder for performing `hard' decoding.

\paragraph*{Soft decoding using IQ data.---}
While ACES provides detailed context-dependent characterisation data of the circuit elements for syndrome extraction, including measurements, it does not leverage the full information available from the measurements used in this superconducting architecture.  We capture the full information available in the IQ values from qubit readout; see Methods Sec.~\ref{sec:IQdata}.  From the IQ data, we can assign probabilities to the measurement outcome being $\ket{0}$, $\ket{1}$, or the leaked state $\ket{2}$.  These probabilities, in turn, can be used for soft decoding.
Such soft information has previously been used to drive neural networks~\cite{Varbonov2025,Bausch2024} and to decode a bit-flip code~\cite{Ali2024}.  Here, we construct an offline decoding graph following Ref.~\cite{pattison2021improvedquantumerrorcorrection}, and then apply soft decoding online on a shot-by-shot basis as detailed in the Methods Sec.~\ref{sec:soft_decoding}.

The IQ data also allows for leakage post-selection as an element on top of soft decoding.  
A decoding round can be rejected if any measurement in that shot has a leakage probability exceeding some cutoff.
Logical error rates after leakage post-selection are then computed conditionally on the subset of accepted decoding rounds.  Details are found in Methods Secs.~\ref{sec:IQdata} and \ref{sec:soft_decoding}.

As a reference point for our investigation, we consider two baselines derived from the calibration snapshot that IBM provides for the devices \footnote{https://quantum.cloud.ibm.com/docs/en/guides/calibration-jobs}, collected on the same day as our experiments. 
The snapshot reports observed error probabilities for gates and measurements on each qubit and two-qubit gate supported by the device. 
The first baseline is constructed by averaging these values over all qubits and two-qubit gates, yielding a single probability per operation type, which amounts to a slightly device-specific version of the standard uniform circuit-level noise model. 
The second baseline, in contrast, directly uses the per-qubit and per-gate error rates, thereby capturing the spatial inhomogeneity of the device.
See Methods Sec.~\ref{sec:calibration_snapshot} for further detail.

\subsection{Analysis of results from memory experiments}

We perform repeated rounds of quantum error correction using the distance-5 dynamic compass code, measuring memory survival rates and inferring logical error rates per round.
We provide two baselines where decoding is based on IBM global or per-qubit calibration data, and compare these results with informed decoding that uses ACES characterisation and soft information from measurements; see Fig.~\ref{fig:ler_analysis_results}.  All decoding is performed using BeliefMatching; although this decoder is inefficient compared with other matching decoders (such as PyMatching), it has the best performance and is suitable for small codes such as this distance-5 instance.

Overall, the improvements relative to the global calibration baseline are substantial: per-round logical error rates decrease by 24.2\% in the $Z$ basis and 38.3\% in the $X$ basis. 
Comparable gains are observed across a range of related experiments presented in the Appendix, in which we consider alternative code placements on the device as well as the distance-5 heavy-hex code; across these experiments, the relative improvements span 21.4\%--25.4\% in the $Z$ basis and 28.2\%--39.4\% in the $X$ basis.  Also shown in the Appendix is a similar analysis for the heavy-hex code with the same placement on the device, for comparison; see Appendix~\ref{app:alternative_setting_results}.

Several aspects of these improvements are noteworthy. 
First, using individual qubit calibration data already yields some improvement, but as expected the full ACES characterisation data provides substantially larger gains. 
This is unsurprising, as ACES produces context-dependent error models derived from characterisation of the syndrome extraction circuits themselves. 
The improvements are, however, strongly basis-dependent: relative to the calibration-based decoder, ACES achieves a 21.0\% relative improvement in the $X$-basis memory experiments, whereas the corresponding $Z$-basis improvement is only 3.0\% and is not statistically significant.  

Second, we do not observe a statistically significant improvement by using a soft decoder without post-selection, for any of the error models.  Details are provided in the Appendix~\ref{app:soft_decoding_analysis}.
However, when rounds are post-selected based on the detection of leakage, we observe significant improvements in logical error rates beyond what was achieved with ACES characterisation data alone, even with a very low per-round abort rate of a few percent.  These gains are most evident in the $Z$-basis memory experiments (a 16.3\% improvement in logical error rate), with $X$-basis memory experiments demonstrating a meaningful but less pronounced improvement (11.3\%).  While these gains arise from a very modest amount of post-selection, we find that it is not possible to gain much further by increasing the amount of post-selection; see Fig.~\ref{fig:ler_analysis_results}(d).  
This indicates that with a small amount of post-selection, leakage is no longer the dominant source of error.

\section{Discussion}

A fault-tolerant quantum computer will require very well-controlled logical qubits with error rates $\sim 10^{-10}-10^{-12}$. To reach this ambitious goal, we will require theoretical innovations that complement and mitigate the ongoing development of experimental hardware. Our results to this end are twofold.

First, we have proposed and tested a code with a dynamic syndrome readout circuit that is readily compatible with the layout and connectivity of a superconducting processor design that has been built and continuously reiterated over several years. Our dynamic compass code has a relatively small footprint, demonstrates a threshold, and has a low-depth circuit that completes a full syndrome extraction cycle with only two sequential rounds of mid-circuit measurements. It therefore offers a promising path to large-scale quantum computing with further development of the heavy-hex architecture. Indeed, in our accompanying paper, we describe large-scale fault-tolerant logic gates based on this code and forecast its scalability using numerical simulations~\cite{DCCnumerics}.

Second, using a distance-5 instance of the code implemented in an experiment, we have demonstrated that logical error rates can be improved by 24\%--38\%, incorporating the detailed and contextual error information from efficient, large-scale characterisation tools into our decoding algorithm. We have shown we can enhance the performance of state-of-the-art matching decoders on real hardware by supplying noise characterisation data from ACES, and by furnishing it with soft information that is obtained from individual shot-by-shot readout operations. Our noise characterisation tools may be readily applied to decoding algorithms that operate on larger-scale codes. As such, it is reasonable to expect that the improvements we have demonstrated here are accessible with these codes. It will be interesting to determine to what extent these tools improve codes in the future as physical qubits develop and code distances become larger.

Looking ahead, it is valuable to interrogate the bottlenecks that have prevented us from obtaining better logical error rates. First, we found that a significant improvement in logical error rate was enabled by post-selecting QEC rounds according to the likelihood that a qubit has experienced leakage. However, post selection in this way is not a scalable approach to fault-tolerant quantum computing since the proportion of rejected shots will increase with circuit depth. Nevertheless, this result is indicative that, although leakage rates are small on these devices, a sizeable part of the logical error budget is due to leakage. Overcoming this issue in a scalable way could be enabled by a mechanism to reset qubits that have undergone leakage, an approach that has been taken in several recent experiments~\cite{Miao2023,Lacroix2025}. Second, an issue we face is inhomogeneities in the qubit lattice, where a small number of qubits and couplers perform substantially worse than others on the device. These defective components significantly impact the logical error rate and therefore make it very difficult to place a high-fidelity code on the available lattice. Indeed, we have presented the best results we could obtain with a distance-5 code, whereas preliminary results attempting to implement a distance-7 code on the device performed poorly due to a small number of qubits with reduced operational fidelities. We expect that this problem will be alleviated as fabrication of quantum devices continues to improve. Naturally, we should certainly not expect a very large scale device to be free from defective qubits. To address this problem for both near-term devices as well as large-scale quantum hardware of the future, moving forward it will be valuable to investigate strategies to mitigate qubit and coupler dropout~\cite{Stace2009thresholds, Auger2017, Strikis2021, Siegel2023adaptivesurfacecode, Lin2024codesign, McLauchlan2024accommodating, Wei2025low, Leroux2025snakes, Debroy2025luciinsurfacecode,  wolanski2026automated, wei2026adaptivedeformationcolorcode,higgott2025handlingfabricationdefectshexgrid}.

\section{Methods}
\subsection{IQ data characterisation}
\label{sec:IQdata}
Qubit measurement on \texttt{ibm\_pittsburgh} is performed via dispersively coupled readout resonators~\cite{Bronn2017}.
The signal from these resonators consists of two components of information, I and Q, corresponding to two orthogonal quadratures of the resonator, which can be used to determine the state of the transmon qubit.
Different regions in IQ phase space correspond to different energy levels of the transmon.
A binary output can be extracted by determining whether a measurement's IQ point lies closer to the $\ket{0}$ region or the $\ket{1}$ region in phase space.
Alternatively, the full values of I and Q can be used to determine a probability that the measured qubit was in either state, and this soft information can be passed to the decoder to potentially improve performance.
Furthermore, since leakage states occupy different regions in phase space, a leakage event can be detected and this information may be used for post-selection.

\begin{figure*}[t]
    \centering
    \includegraphics[width=\linewidth]{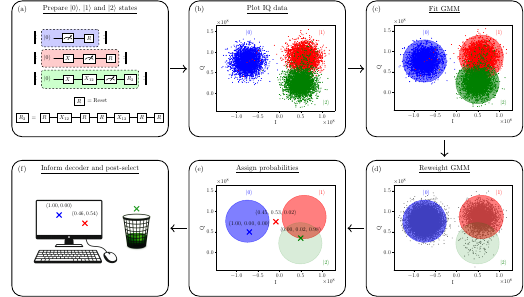}
    \subfloat{\label{fig:IQ-data-circuit-diagrams}}%
    \subfloat{\label{fig:IQ-data-callibration-data}}%
    \subfloat{\label{fig:IQ-data-fit-GMM}}%
    \subfloat{\label{fig:IQ-data-reweight-GMM}}%
    \subfloat{\label{fig:IQ-data-assign-probs}}%
    \subfloat{\label{fig:IQ-data-inform-decoder}}%
    \caption{\textbf{IQ data characterisation pipeline.} 
    (a) Circuit diagrams for preparing the $\ket{0}$, $\ket{1}$ and $\ket{2}$ states of a transmon. 
    $R$ denotes the reset operation and $R_2$ denotes a sequence of operations designed to remove leakage.
    Each circuit is repeated $4000$ times.
    (b) IQ data is plotted for each measurement.
    Each point has been colour coded by its corresponding circuit in (a).
    (c) A Gaussian Mixture Model (GMM) is fitted to the calibration data.
    The ellipse perimeters represent 3 standard deviations.
    (d) The GMM is reweighted using the data from the error correction experiment.
    The relative intensities of the ellipses indicate the relative weights.
    (e) Probabilities are assigned to each measurement from the error correction experiment using the reweighted GMMs.
    (f) The probabilities for each measurement are used to inform the decoder, and shots are discarded if they are deemed to have leaked.
    }
\end{figure*}

In order to obtain state probabilities for each measurement outcome based on the IQ values from readout, we use a procedure similar to that outlined in Ref.~\cite{sundaresan2022}.
First, we obtain calibration data by preparing many instances of the $\ket{0}$, $\ket{1}$ and $\ket{2}$ states of the transmon on each qubit.
\Cref{fig:IQ-data-circuit-diagrams} shows the circuit diagrams for preparing each state.
The $\ket{1}$ state is prepared using an $X$-gate, whereas the $\ket{2}$ state is prepared using an $X$-gate followed by an $X_{12}$ gate. The latter gate transfers population between the $\ket{1}$ and $\ket{2}$ states of the transmon.
After each state is prepared, it is measured using either a longer or shorter measurement pulse (see \cref{sec:calibration_snapshot,sec:experiment-settings}), and then reset.
To reset the qubit after preparing a $\ket{2}$ state, we perform the reset procedure shown at the bottom of \cref{fig:IQ-data-circuit-diagrams}, which involves multiple reset operations interspersed with $X_{12}$ gates to remove leakage.
Each circuit is repeated $4000$ times per state on each qubit for both measurement pulses.
To minimise the chance of residual excitations, we prepare all $\ket{0}$ states first, followed by all $\ket{1}$ states and then all $\ket{2}$ states.

The resulting calibration data for the shorter measurement pulse on Qubit 129 is shown in \cref{fig:IQ-data-callibration-data}.
Using this data, we fit a Gaussian Mixture Model (GMM) whose probability density function is given by
\begin{equation} \label{eqn:GMM}
    g(I,Q) = \sum_{i=0}^2 w_i \mathcal{N}_{\bm{\mu}_i, \Sigma_i}(I,Q),
\end{equation}
where $\mathcal{N}_{\bm{\mu}, \Sigma}$ is a two-dimensional Gaussian with mean $\bm{\mu} = (\mu_I, \mu_Q)$ and covariance matrix $\Sigma$, and $w_i \in [0, 1]$ denote the weights for each Gaussian.
We use the Python package \texttt{scikit-learn} to fit this GMM to the data.
We find that the data is well described by Gaussians with a diagonal covariance matrix, consistent with the fact that the noise in each quadrature should be uncorrelated~\cite{Wiseman1998}, and enforce this in the fitting procedure.
However, the results are negligibly changed if we allow for non-diagonal covariance matrices.
\Cref{fig:IQ-data-fit-GMM} shows the fitted GMMs to the data for Qubit 129, where the perimeter of each ellipse represents three standard deviations.

Using the GMM, an IQ data point can be assigned a three-tuple of probabilities $\bm{p} = (p_0, p_1, p_2)$, where
\begin{equation}
    p_i(I, Q) = \frac{w_i \mathcal{N}_{\bm{\mu}_i, \Sigma_i}(I,Q)}{g(I,Q)}.
\end{equation}
A measurement outcome is then classified as state $\ket{i}$ if $i = \mathrm{argmax}\,\bm{p}$.

Since equal numbers of each state are prepared in the calibration data, the weights $w_i$ in \cref{eqn:GMM} are approximately equal. 
However, in an error-correction circuit, the probability of leakage should be low, meaning that $w_2 \ll w_0, w_1$.
Therefore, we update the weights of the GMM by using the data from each error correction experiment (see \cref{fig:IQ-data-reweight-GMM}).
This is achieved by classifying all measurements from a particular error correction experiment using the GMMs from the calibration data, and then updating the weights in the GMM according to the relative number of each state that has been classified.
This is then repeated until the weights remain stable (we find that it takes $\sim 15$ iterations for the weights to change by less than $10^{-6}$).
Once the weights have stabilised, then the GMM with the updated weights are used to classify the measurements from the error correction experiment (see \cref{fig:IQ-data-assign-probs}), which we describe in the next section.

\subsection{Soft decoding and leakage post-selection pipeline}
\label{sec:soft_decoding}

To decode the experimental data, we first construct an offline decoding graph for each code and basis setting.
Specifically, we generate noisy memory circuits for the heavy-hex and dynamic compass codes using an ACES-derived noise model that includes gate-dependent Pauli channels, idle noise, pre-measurement noise, classical readout error, and post-measurement noise.
From these circuits we derive a detector error model (DEM), which characterises independent error mechanisms and the corresponding set of detectors (i.e., specific products of measurement outcomes that are deterministic when there are no errors) flipped by them.
As in surface-code decoding, every error mechanism in the DEM can be decomposed into ``edge-like'' mechanisms that affect at most two detectors, allowing the use of minimum-weight perfect-matching (MWPM) based decoders such as PyMatching \cite{Higgott2025sparseblossom} and BeliefMatching \cite{higgott2023improved}.
In particular, although some physical error processes, such as $Y$ errors and other correlated errors from two-qubit gates, can flip more than two detectors simultaneously, these higher-weight detection events can be treated as composites of edge-like mechanisms already present in the DEM rather than as fundamentally new hyperedges.
This circuit-to-DEM compilation is carried out once for a given experimental setting and is then reused for all shots.

Soft decoding is then applied online on a shot-by-shot basis.
For each measurement, the probability three-tuple $\bm{p} = (p_0, p_1, p_2)$ yielded by the GMM is used to update the classical readout error rate associated with that measurement in the DEM by the shot-dependent value $\min(p_0, p_1)/(p_0 + p_1)$.
In this way, each shot is decoded using its own shot-specific DEM, while all non-measurement noise parameters remain fixed by the offline ACES characterisation.
Soft decoding may therefore be viewed as directly incorporating measurement uncertainty inferred from the IQ data into the detector-level decoding graph.

Leakage post-selection is implemented as an optional additional analysis layer on top of soft decoding.
For a chosen cutoff probability $p_\mathrm{cutoff}$, a shot is rejected if any measurement in that shot satisfies $p_2 > p_\mathrm{cutoff}$.
Varying $p_\mathrm{cutoff}$ therefore interpolates continuously between stronger leakage suppression (leading to lower logical error rates) and higher sample acceptance, yielding the trade-off shown in \cref{fig:ler_analysis_results}(b).
Logical error rates after leakage post-selection are computed conditionally on the subset of accepted shots.

\subsection{Calibration-snapshot baseline noise models}
\label{sec:calibration_snapshot}

We compare our ACES-derived noise model against two simpler models constructed from the calibration snapshot data by IBM Quantum.
For each device, IBM reports error rates for individual qubits and qubit connections, which are updated on a regular schedule.
We use the calibration snapshot recorded on the same day as our experimental runs.
The snapshot contains the following per-qubit or per-connection error rates:
\begin{itemize}
    \item \textbf{MEASURE} and \textbf{MEASURE\_2}: measurement outcome flip error rates associated with the longer and shorter measurement pulses, respectively:
    \begin{align*}
        \frac{1}{2} [&\Pr(\text{outcome}=1 \mid \text{prepared } \ket{0}) \\
        &+ \Pr(\text{outcome}=0 \mid \text{prepared } \ket{1})].
    \end{align*}
    \item \textbf{ID error}: single-qubit error rate for idling periods during gate layers.
    \item \textbf{RX error}: single-qubit gate error rate for $R_X$ gates. In the calibration data we use, this value is always identical to the ID error for every qubit.
    \item \textbf{CZ error}: two-qubit error rate reported per qubit connection (i.e., per bond rather than per qubit).
\end{itemize}

From this data we construct two baseline noise models by treating \textbf{RX error} and \textbf{CZ error} as single-qubit and two-qubit depolarising noise channels after corresponding ideal gates, \textbf{ID error} as single-qubit depolarising channels on idling qubits for each layer, and \textbf{MEASURE}/\textbf{MEASURE\_2} as bit-flip errors before measurements (i.e., pre-measurement noise).
In the \emph{calibration (global)} model, we average the error rate of each error type over all qubits or connections and apply the resulting uniform rate to every corresponding location in the circuit.
This model may be regarded as a slightly more realistic alternative to a standard depolarising noise model, since the relative magnitudes of single-qubit, two-qubit, and measurement errors reflect the actual device characteristics, even though spatial variation is discarded.
In the \emph{calibration (per-qubit)} model, each qubit or connection retains its individually reported error rate, thereby capturing the spatial inhomogeneity of the device noise.
When a qubit or bond is absent from the snapshot, the missing value is replaced by the average over all reported values of the same error type, analogous to the global model.

Both models involve several simplifying assumptions.
First, the calibration snapshot does not decompose measurement outcome flip errors into separate pre-measurement quantum noise and classical readout error; we thus modelled them as pre-measurement noise only.
Second, post-measurement noise is not included in the snapshot data, as it does not flip the measurement outcome but just changes the post-measurement state, thus we do not include it in the noise models as well.
Second, because measurements take substantially longer than gates ($\approx$\ 2.6 $\mu$s .v. 88 ns), qubits that idle during a measurement layer may experience higher noise than the reported ID error rate, which is ignored in our models.

\subsection{Experiment settings}
\label{sec:experiment-settings}

The results in \cref{sec:results} use data from experimental runs performed on \texttt{ibm\_pittsburgh} on 24 February 2026.
We study distance-5 implementations of the heavy-hex and dynamic compass codes, with Qubit 46 placed at the top-left corner of the code placement.
(We also acquired data for the Qubit 65 placement of this code.  Results are presented in Appendix~\ref{app:alternative_setting_results}.)
For each code, we analyse preparations in the four logical basis states $\ket{0}$, $\ket{1}$, $\ket{+}$ and $\ket{-}$ over numbers of syndrome-extraction rounds $T \in \{0,1,2,4,6,8,10,12,15,20,30\}$.
Final data-qubit measurements use the longer measurement pulse, whereas mid-circuit syndrome measurements use the shorter pulse.
We therefore use separate GMM calibrations for these two classes of measurements.

\subsection{Performance evaluation}

To obtain performance metrics that are independent of the number of rounds $T$, we compute the logical error rate $p_\mathrm{L}(T)$ and acceptance probability $p_\mathrm{acc}(T)$ as functions of $T$ and convert them into per-round rates.
We fit $\log(1 - 2 p_\mathrm{L}(T))$ linearly as a function of $T$, and use the fitted slope $G$ to define the per-round logical error rate $\epsilon_\mathrm{L} = (1 - e^{G})/2$.
This form follows from the standard phenomenological model in which each round produces an independent logical error with probability $\epsilon_\mathrm{L}$, and an overall logical error occurs when an odd number of rounds fail~\cite{obrien2017density, GoogleQuantumAI-ExponentialSuppressionBit-2021}.
Similarly, we fit $\log p_\mathrm{acc}(T)$ linearly as a function of $T$, and use the fitted slope $c$ to define the per-round abort rate $\epsilon_\mathrm{abort} = 1 - e^{c}$, assuming that each round contributes an approximately constant multiplicative acceptance factor $1 - \epsilon_\mathrm{abort}$.

%

\section{Acknowledgments}
We thank Andrew Doherty for useful discussions about the interpretation of IQ data, and we thank the organisers of the Quantum Error Correction Workshop 2025 at the Yukawa Institute for Theoretical Physics where the initial ideas for this project were conceived.
BJB is grateful for the hospitality of the Center for Quantum Devices at the University of Copenhagen. We acknowledge support from the Intelligence Advanced Research Projects Activity (IARPA), under the Entangled Logical Qubits program through Cooperative Agreement Number W911NF-23-2-0223 (R.H., B.J.B., and S.D.B.).  The views and conclusions contained in this document are those of the authors and should not be interpreted as representing the official policies, either expressed or simplied, of IARPA, the Army Research Office, or the U.S. Government. The U.S. Government is authorized to reproduce and distribute reprints for Government purposes notwithstanding any copyright notation herein. S.H.L was supported by the National Research Foundation of Korea (NRF) grant (RS-2026-25476454, RS-2024-00442710, RS-2023-NR068116) funded by the Ministry of Science and ICT (MSIT) of the Korean government. J.Z. acknowledges the support of the 2025 Google PhD Fellowship. This work was supported in part by Japan Science and Technology Agency (JST) as part of Adopting Sustainable Partnerships for Innovative Research Ecosystem (ASPIRE), Grant Number JPMJAP25A3.

\appendix

\widetext

\section{Experimental results on the heavy-hex code and an alternative code placement \label{app:alternative_setting_results}}

\begin{figure*}[h!]
    \includegraphics[width=\linewidth]{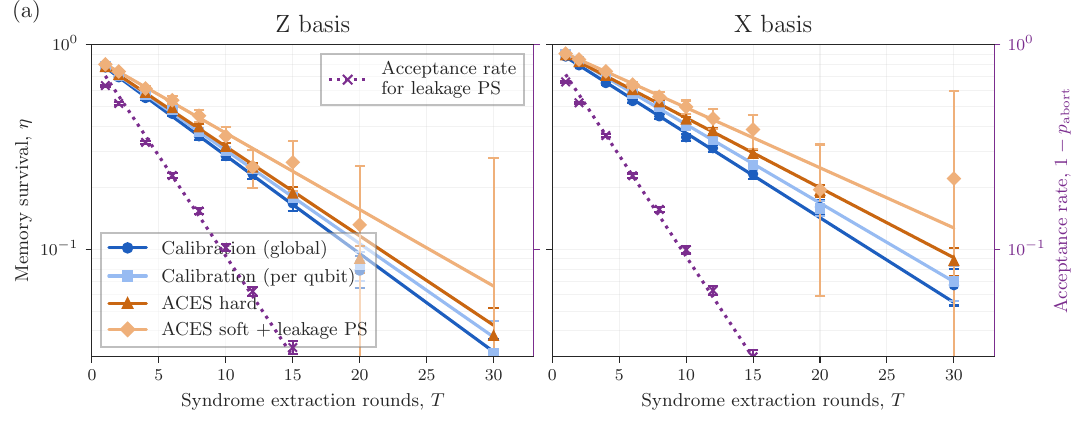} \\
    \includegraphics[width=0.6\linewidth]{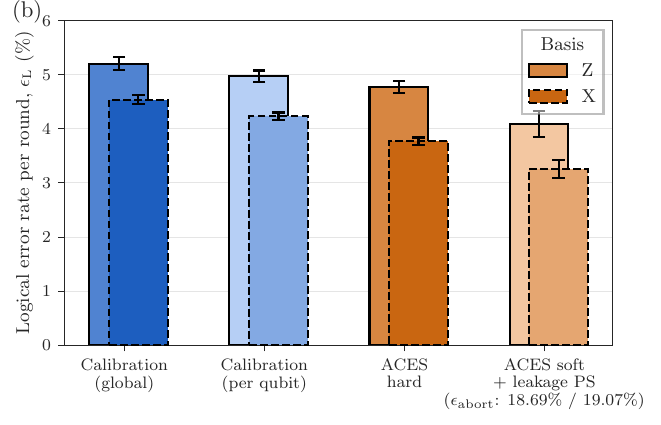}
    \includegraphics[width=0.39\linewidth]{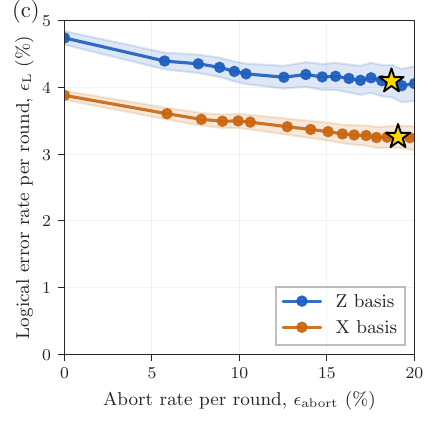}
    \caption{Memory experiment results for the distance-5 dynamic compass code at the alternative placement (top left qubit~65).}
\label{fig:ler_analysis_results_dcc_q65_appendix}
\end{figure*}

\begin{figure*}[h!]
    \includegraphics[width=\linewidth]{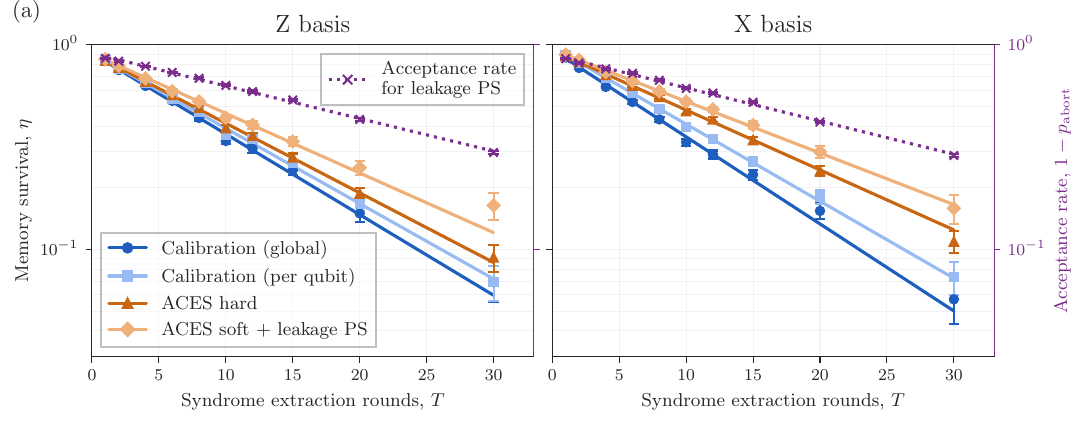} \\
    \includegraphics[width=0.6\linewidth]{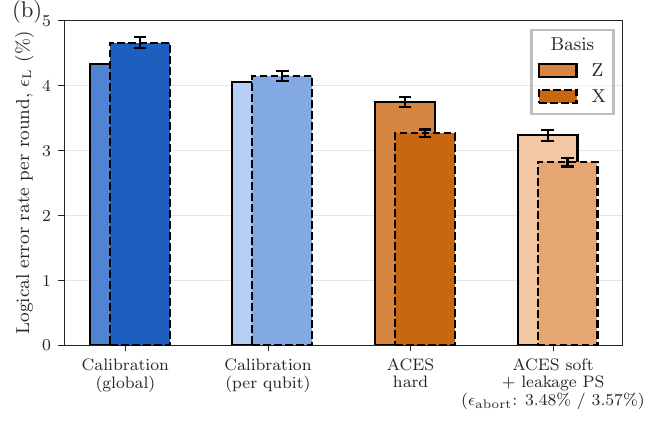}
    \includegraphics[width=0.39\linewidth]{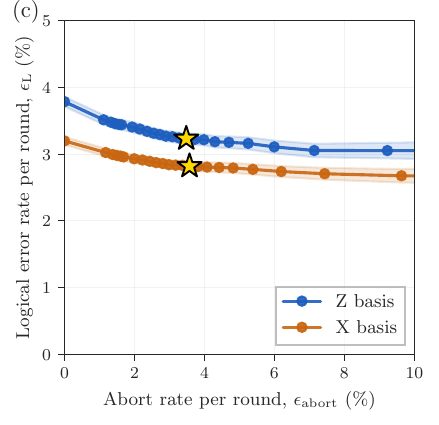}
    \caption{Memory experiment results for the distance-5 heavy-hex code at the default placement (top left qubit~46).}
\label{fig:ler_analysis_results_heavyhex_q46_appendix}
\end{figure*}

\begin{figure*}[h!]
    \includegraphics[width=\linewidth]{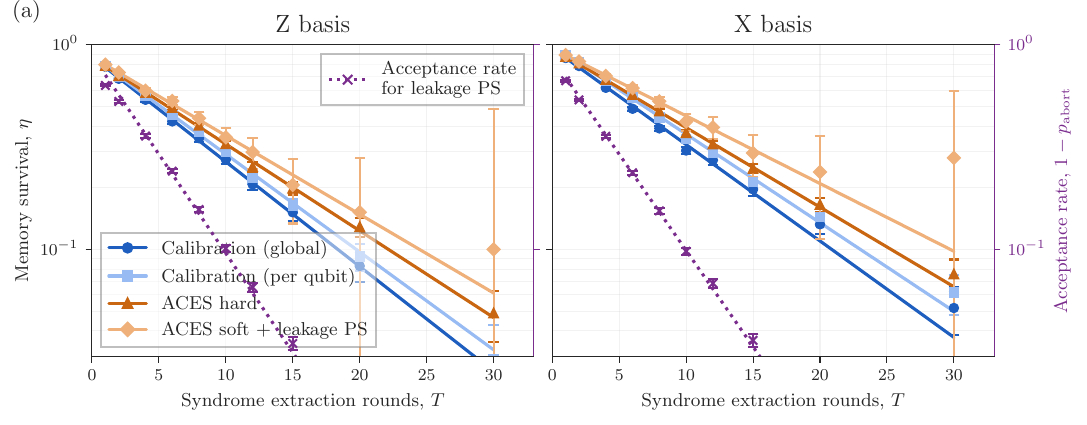} \\
    \includegraphics[width=0.6\linewidth]{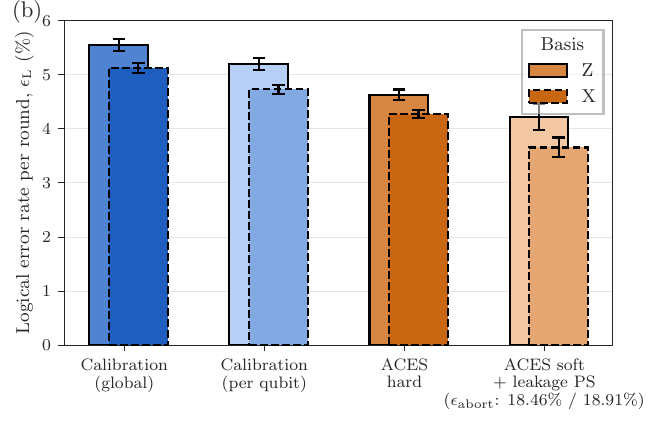}
    \includegraphics[width=0.39\linewidth]{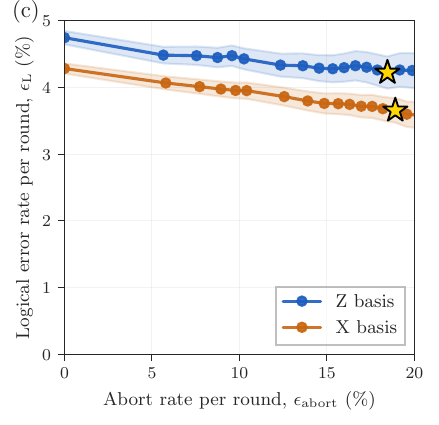}
    \caption{Memory experiment results for the distance-5 heavy-hex code at the alternative placement (top left qubit~65).}
\label{fig:ler_analysis_results_heavyhex_q65_appendix}
\end{figure*}

In the main text, we presented experimental results for the distance-5 dynamic compass code (DCC) instance with the specific placement on the \texttt{ibm\_pittsburgh} processor as shown in Fig.~\ref{fig:device_schedule}, with Qubit~46 (q46) as the top-left qubit.
For comparison, we have also taken data for the distance-5 heavy-hex code~\cite{chamberland2020, harper2025characterising} based on the same footprint, and for an alternative code placement with Qubit~65 (q65) as the top-left qubit.
Figures~\ref{fig:ler_analysis_results_dcc_q65_appendix}--\ref{fig:ler_analysis_results_heavyhex_q65_appendix} present the results for (DCC, q65), (heavy-hex, q46), and (heavy-hex, q65), respectively.
Across all configurations, our strategies combining ACES-derived noise models with leakage post-selection yield substantial gains over global decoding, reducing the logical error rate by 21.4\%--25.4\% in the $Z$ basis and 28.2\%--39.4\% in the $X$ basis.

Comparing the q46 and q65 placements, the latter exhibits 5--6 times higher abort rates at a fixed leakage-probability cutoff of 0.5.
This indicates that there are one or more qubits that are more susceptible to leakage in the q65 placement than in the q46 placement.
Specifically, when we compare the leakage-cluster weights of the syndrome qubits unique to each placement, the q46-only and q65-only sets have mean weights of 0.0061 and 0.032 and maximum weights of 0.026 and 0.13, respectively; the q65-only syndrome qubits therefore exhibit roughly five times higher leakage probabilities on average.

Comparing the two codes, the heavy-hex code attains a lower $Z$-basis error rate but a higher $X$-basis error rate than the DCC.
This is expected because, unlike the DCC, the heavy-hex code has no threshold in the $X$ basis owing to its high-weight $X$-type stabilisers, while $Z$-type stabilisers are measured more frequently than in the DCC.
This behaviour is consistent with the numerical results reported in our accompanying paper~\cite{DCCnumerics}.

\clearpage

\section{Detailed analysis of soft information decoding \label{app:soft_decoding_analysis}}

\begin{figure*}[tb!]
    \includegraphics[width=\linewidth]{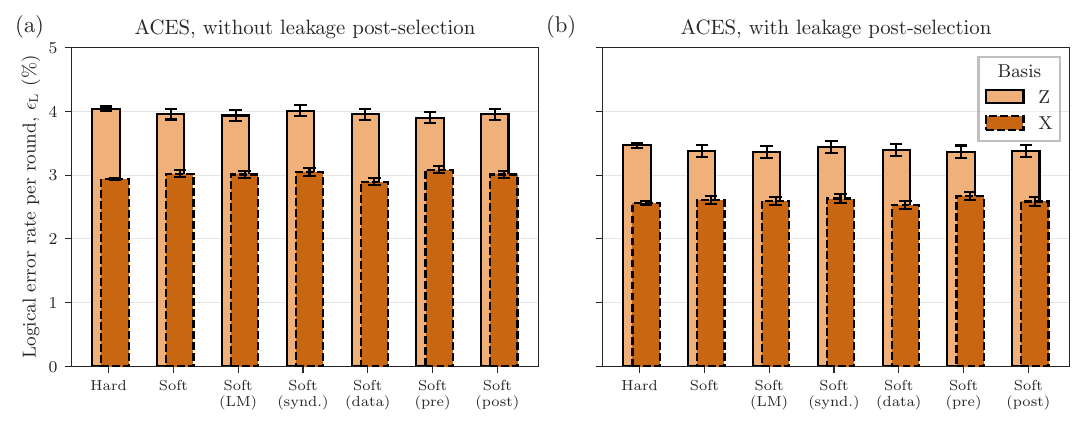}
    \caption{\textbf{Comparison of soft-information decoding strategies for the distance-5 dynamic compass code.}
    Per-round logical error rates obtained with hard decoding and with the six soft-decoding variants defined in the text, (a)~with and (b)~without leakage post-selection.
    In all cases, the underlying detector error model is derived from the ACES noise characterisation, and leakage post-selection, where applied, uses a cutoff of $p_\mathrm{cutoff} = 0.5$.
    }
    \label{fig:ler_bar_plot_q46_soft_decoding_comparison}
\end{figure*}

As noted in the main text, the soft information extracted from the IQ readout is useful for leakage post-selection, where it yields substantial reductions in the per-round logical error rate of 11.3\%--16.3\% relative to hard decoding.
In contrast, we do not observe statistically significant improvements from soft decoding alone, without leakage post-selection, in our memory experiments.
To better understand this behaviour, in this appendix we examine several soft-decoding variants motivated by distinct physical considerations and compare their performance with the hard-decoding baseline.

\Cref{fig:ler_bar_plot_q46_soft_decoding_comparison} summarises the comparison.
In each variant, the probability three-tuple $\bm{p} = (p_0, p_1, p_2)$ yielded by the GMM (see Methods Secs.~\ref{sec:IQdata} and \ref{sec:soft_decoding}) is used to update measurement-related error rates in the DEM on a shot-by-shot basis, while all non-measurement noise parameters are kept fixed to the values obtained from the offline ACES characterisation.
The variants differ only in the precise update rule and in the subset of measurements to which it is applied:
\begin{itemize}
    \item \textbf{Soft}: the default scheme introduced in Methods Sec.~\ref{sec:soft_decoding}.
    For every measurement, the classical readout error rate in the DEM is updated to $\min(p_0, p_1)/(p_0 + p_1)$.

    \item \textbf{Soft (leakage-marginalized; LM)}: identical to \textbf{Soft}, except that the shot-dependent measurement error rate is instead set to $\min(p_0, p_1) + p_2/2$.
    This choice is motivated by the intuition that a leakage event makes the measurement outcome bit maximally uncertain and therefore contributes $p_2/2$ to the effective flip probability.

    \item \textbf{Soft (synd.)}: identical to \textbf{Soft}, but the update is applied only to the mid-circuit measurements of the syndrome qubits; the final data-qubit measurements are left unchanged.

    \item \textbf{Soft (data)}: identical to \textbf{Soft}, but the update is applied only to the final data-qubit measurements; the mid-circuit syndrome measurements are left unchanged.

    \item \textbf{Soft (pre)}: the shot-dependent measurement error rate is partitioned between the classical readout error and the pre-measurement bit-flip error in the DEM, rather than being attributed entirely to the former.
    The partition ratio is fixed to match the corresponding ratio in the ACES-derived noise model.
    This variant is motivated by the fact that our circuits for fitting GMMs in Fig.~\ref{fig:IQ-data-circuit-diagrams} cannot intrinsically distinguish classical readout errors from pre-measurement bit-flip errors.

    \item \textbf{Soft (post)}: identical to \textbf{Soft}, but the post-measurement bit-flip error rate in the DEM is additionally updated, on a shot-by-shot basis, to $2 p_1 p_\mathrm{post}^\mathrm{ACES}/(p_0 + p_1)$, where $p_\mathrm{post}^\mathrm{ACES}$ denotes the post-measurement bit-flip error rate reported by ACES.
    This form is motivated by the expectation that post-measurement bit flips are dominated by relaxation from the $\ket{1}$ state (so the update scales with $p_1$), and it is constructed so that it reduces to $p_\mathrm{post}^\mathrm{ACES}$ when $p_0 = p_1$.
\end{itemize}

Despite being motivated by these distinct physical considerations, none of the variants yield statistically significant improvement compared to the hard-decoding baseline. 
Notably, \textbf{Soft (data)}, which updates only the final data-qubit measurements, performs comparably to \textbf{Soft} in all cases. 
This pattern can be interpreted in terms of the space-time structure of the decoding graph.
Measurement errors on syndrome qubits produce timelike detection events, whereas errors on the final data-qubit measurements produce spacelike detection events that can directly contribute to the logical failure of a memory experiment.
From this perspective, it is not particularly surprising that soft information on the syndrome measurements provides little additional benefit.
In contrast, soft information on the final data-qubit measurements may directly contribute to reducing spacelike errors, which is consistent with the favourable performance of \textbf{Soft (data)}.
This raises the possibility that utilising soft information from syndrome measurements for decoding could be more effective in stability experiments~\cite{Gidney2022dual,harper2025characterising}, which measure the ability for a QEC code to correct timelike errors, and we believe this is worth exploring in future work


Additionally, a complementary interpretation is provided by the Chernoff-information framework for soft readout decoding~\cite{danjou2021generalized}. 
For repeated quantum nondemolition (QND) readout of a binary observable, the advantage of retaining analog readout information can be quantified by
$A=C/C_b$, where $C$ is the Chernoff information of the analog readout distributions and $C_b$ is the Chernoff information after hard binarisation.
When $A\simeq 1$, analog-to-binary conversion discards little information relevant for distinguishing the two binary outcomes, and little improvement from soft decoding is expected.
Ref.~\cite{danjou2021generalized} further shows that this situation can arise when readout errors are dominated by conversion-like events rather than by Gaussian overlap near the decision boundary, whereas purely Gaussian readout noise would generally lead to a nontrivial soft-decoding advantage.
Our results are consistent with this perspective: the IQ-derived soft decoder used here updates only shot-dependent classical measurement error rates in the DEM, while all non-measurement noise parameters are fixed by the ACES characterisation.
Thus, if the dominant logical failures are not caused by ambiguous binary readout outcomes, or if the most useful information in the IQ record is leakage-related rather than $0/1$-confidence-related, then little improvement from soft decoding alone is expected.
We emphasise, however, that we do not compute $C/C_b$ for our IQ distributions here, so this should be viewed as a qualitative interpretation rather than a direct prediction.

\end{document}